
\input epsf

\ifx\epsffile\undefined\message{(FIGURES WILL BE IGNORED)}
\def\insertfig#1#2{}
\else\message{(FIGURES WILL BE INCLUDED)}
\def\insertfig#1#2{{{\baselineskip=4pt
\midinsert\centerline{\epsfxsize=3.5truein\epsffile{#2}}{{\centerline{#1}}}
\medskip\endinsert}}}
\def\insertfigbig#1#2{{{\baselineskip=4pt
\midinsert\centerline{\epsfxsize=5.5truein\epsffile{#2}}{{\centerline{#1}}}
\medskip\endinsert}}}
\def\insertfigsmall#1#2{{{\baselineskip=4pt
\midinsert\centerline{\epsfxsize=2.5truein\epsffile{#2}}{{\centerline{#1}}}
\medskip\endinsert}}}

\fi

\input harvmac
\input ucsdmac

\def\hbar{\bar h_Q}

\def\qsl{\hbox{/\kern-.5600em {$q$}}}
\def\ksl{\hbox{/\kern-.5600em {$k$}}}

\def\({\left(}
\def\){\right)}

\def\OMIT#1{}
\def\frac#1#2{{#1\over#2}}

\def\vbc{$|V_{cb}|$}

\hbadness=10000

\noblackbox
\vskip 2.in
\centerline{{\titlefont{HQET and Inclusive Decays of Heavy Hadrons}
\footnote{*}{{\tenrm Work
supported in part by the Department of Energy under contract
DE--FG02--91ER40682.}}}}
\vskip 0.5in
\centerline{Martin J.~Savage}
\medskip
{\it{
\centerline{Department of Physics, Carnegie Mellon University,
Pittsburgh PA 15213}}}

\vskip 1.2in

\abstract{
We discuss the recent developments in inclusive
decays of hadrons containing a heavy quark.
The subject is approached in  a model independent
way using an operator  product expansion
for the time-ordered product of weak currents.
We discuss the extraction of the weak mixing
angle $|V_{cb}|$ from inclusive
semileptonic $B$ decays and the application of these
techniques to the  endpoint of the $e^-$ spectrum in
$B\rightarrow X_ue\nu$.
We also suggest that the comparison of inclusive semileptonic
widths for  $D_s$ and $D$
mesons will give an indication of the consistency of this
approach in  charm decays.
\bigskip
{\it Invited talk presented at the "Intersections of Particle and Nuclear
Physics" conference, St Petersburg, Florida, 31 May to 6 June, 1994.}
}

\vfill
\UCSD{\vbox{
\hbox{CMU-HEP 94-19}
\hbox{DOE-ER/40682-73}}
}{June 1994}
\eject

The most exciting recent developments in the theory of hadrons containing
heavy quarks have been in the area of inclusive decays and it is this subject I
will
be reviewing.
I am assuming that the audience has some familiarity with the
subject of heavy  quark physics and
the basic elements of HQET, although I will not be  drawing on it much.
Let me begin by reminding you of the reasons for examining decays
of heavy hadrons,  besides their intrinsic interest from the standpoint of pure
QCD.

The lagrange density describing the weak charged current
interactions is
\eqn\weak{{\cal L}_{\rm cc}\  \ =\ \  V_{ij}\ {ig_2\over 2\sqrt{2}}\
\overline{u}^i\gamma^\mu (1-\gamma_5) d^j\  W^-_\mu\ \ \ +\ \
{\rm h.c}\ \ \ \ ,}
where $g_2$ is the $SU(2)_L$ coupling constant and
$V_{ij}$ is a complex number with magnitude less than
unity appropriate for the $j\rightarrow i$ transition.
The nine different $V_{ij}$'s  form the
Cabibbo-Kobayashi-Maskawa (CKM) mixing matrix described by
three angles $\theta_{1,2,3}$  and one phase $\delta$.
The unitarity of the CKM matrix gives rise to relations between
elements that can be conveniently displayed in the complex plane, eg.,
\eqn\ckm{ V^*_{ud}V_{ub}+V^*_{cd}V_{cb}+V^*_{td}V_{tb} = 0 \ \ \ .}
This forms a triangle in the complex plane whose area and
lengths of  sides are invariant under phase redefinitions of the quark fields,
\fig\one{do do do}.

\insertfig{Fig~1. The unitary triangle.  The phase convention is such that
$V^*_{cb}V_{cd}$ is real.}{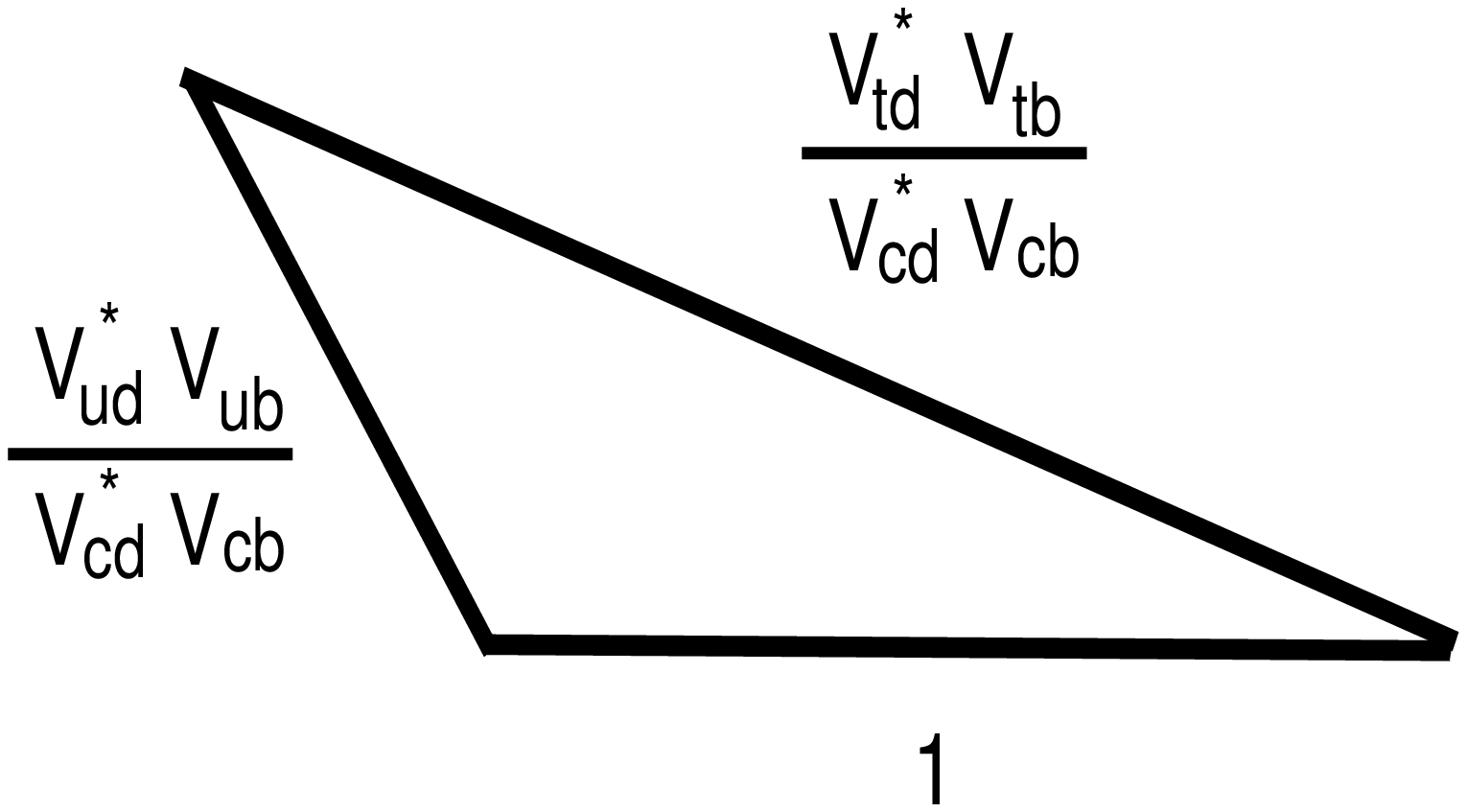}

Any CP violation resulting from CKM mixing is proportional to the area of this
triangle.
It is crucial to determine the lengths of the sides of this triangle as
precisely as
possible {\it independently } of CP violating observables in order to
determine the consistency of the CKM description of CP violation.
If in fact, it is determined that
the lengths of the sides are consistent with a triangle of vanishing
area, then the CP violation  observed in the kaon
sector must arise from physics beyond the  standard model.
This is just one example of the  implications of precise measurements on
the unitarity triangle. In this talk we will be adressing the extraction of
\vbc\
from inclusive decays of hadrons containing a $b$-quark.
The present data set of exclusive decays $B\rightarrow D^{(*)}e\nu$ gives
a value  of \vbc $=0.039\pm 0.05$
\ref\stone{S. Stone, {\it Semileptonic B Decays }, to appear in {\bf B
Decays}, 2nd. Edition,
ed. S. Stone, World Scientific, Singapore.} .
It is possible that inclusive decays of $B$ mesons will allow us to reduce the
uncertainty
in this number.

The definition of the semileptonic inclusive decay width of a b-hadron is
the width for a hadron containing a b-quark to decay semileptonically into a
final state
containing an
$e^-\overline{\nu}_e$ pair and any hadrons, as depicted in Fig.~2.

\insertfig{Fig~2. A cartoon of inclusive decays. }{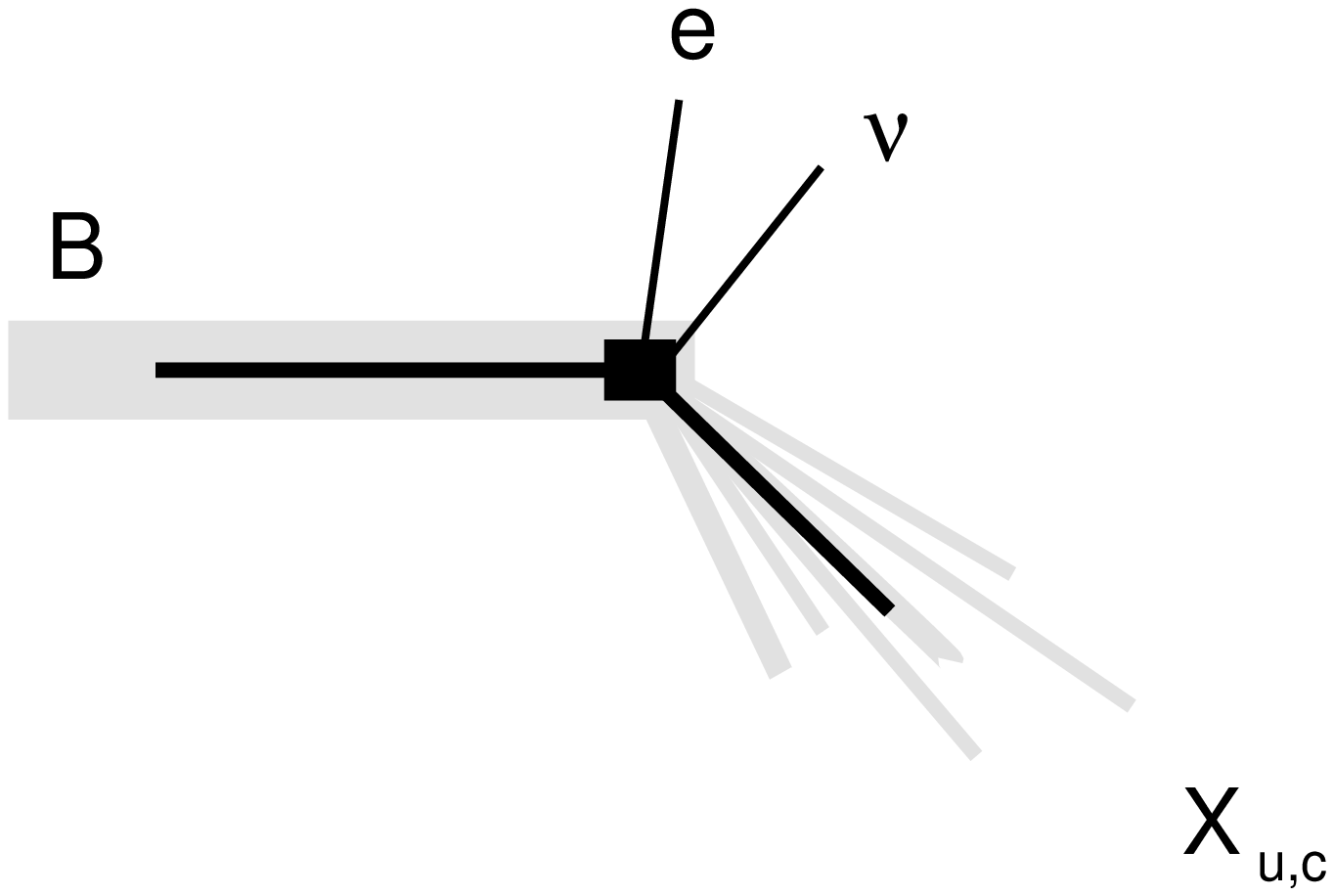}

Our understanding of QCD at short distances suggests that the (inclusive) weak
decay of a hadron  containing a heavy quark
Q with mass infinitely greater than  $\Lambda_{\rm QCD}$ be described by the
weak
decay of the heavy quark  itself. This is a
highly nontrivial statement.  Not only do we expect that the total decay
rate for the hadron be reproduced by the decay rate of the heavy quark (global
duality) but also that the dalitz plot for the decay (after suitable averaging)
be
reproduced by the dalitz
plot obtained from the decay of the heavy quark  (local duality).
Until recently, it had been {\it accepted} that this parton model
description was correct,
however, there was no consistent framework in which this result was
derivable and in
particular there was no framework in which to systematically compute
corrections arising
from the nonzero value of $\Lambda_{\rm QCD}/M_Q$ for the $c$ and $b$
quarks. It was shown by Chay, Georgi and Grinstein in 1990
\ref\cgg{J. Chay, H. Georgi and B. Grinstein, \pl{247}{1990}{399}.}
that these inclusive decays could be handled in ways similar to that used
in deep-inelastic scattering or $e^+e^-$ annihilation.   This method has
allowed the
computation of corrections arising from the finite heavy quark mass up to terms
of
order  $\left( \Lambda_{\rm QCD}/M_Q\right)^2$
\ref\bsuva{I.I. Bigi, M. Shifman, N.G. Uraltsev and A.I. Vainshtein,
\prl{71}{1993}{496}.}\nref\mwa{A.V. Manohar and M.B. Wise,
\physrev{49}{1994}{1310}.}\nref\mannel{T. Mannel,
\np{413}{1994}{396}.}\nref\fls{A.F. Falk, M. Luke and M.J. Savage,
\physrev{49}{1994}{3367}.}\nref\buv{I.I.
Bigi, N.G. Uraltsev and A.I.Vainshtein, \pl{293}{1992}{430}.}-\ref\bksv{B.
Blok, L. Koyrakh, M.
Shifman and A. Vainstein, NSF-ITP 93-68, hep-ph/9307247.}.
At this point it is worth pointing out that the same physics applies to
other processes that are
factorisable such as radiative decays, $B\rightarrow X_s\gamma$ and
lepton pair decays,   $B\rightarrow X_sl^+l^-$.
It has also been suggested that this same procedure can be used to treat
nonleptonic inclusive decays of hadrons containing heavy quarks, see eg.
\ref\bbsuv{I.I. Bigi, B. Blok, M. Shifman, N.G. Uraltsev and A.  Vainshtein,
cern-th-7132-94, (1994).}.
It is not at all clear that this is justified for $b$ and $c$ quarks and
we will not discuss  it further in this talk.

The semileptonic inclusive decay rate for a $B$ meson is given by
\eqn\semia{\eqalign{
 \Gamma & (B\rightarrow e^-\overline{\nu}_e+ anything ) =
|V_{qb}|^2 {G_F^2\over M_B}  \int  {{\rm d}^3 P_e\over (2\pi)^3 2 E_e}
{{\rm d}^3 P_\nu\over (2\pi)^3 2 E_\nu} L^{\mu\nu} \cr
&
\sum_{X_q} {{\rm d}^3 P_X\over (2\pi)^3 2 E_X}
\langle B | j^\dagger_\mu | X_q \rangle \langle X_q |j_\nu | B \rangle
(2\pi)^4 \delta^4 (P_B-P_X-P_e-P_\nu ) }\ \ \ ,}
where $j_\mu =  {1\over 2}\overline{c}\gamma_\mu (1-\gamma_5) b$
( for other factorisable processes we can generalise this to
$j =  \overline{c}\Gamma b$)
and  $L^{\mu\nu}$ is the leptonic tensor
$L^{ab} = 4( P_e^aP_\nu^b + P_e^bP_\nu^a-g^{ab}P_e\cdot P_\nu )$.
The sum over all possible final states allows us to  relate the width to
an integral over the time-ordered product of weak currents, as
\eqn\semib{\eqalign{ \Gamma (B\rightarrow e^-\overline{\nu}_e+ anything ) = &
|V_{qb}|^2 {G_F^2\over 2M_B} \int {{\rm d}^3 P_e\over (2\pi)^3 2 E_e} {{\rm
d}^3
P_\nu\over (2\pi)^3 2  E_\nu} L^{\mu\nu}  \cr
&{\rm Im}\left(-\langle B |T( j^\dagger_\mu\  j_\nu ) | B \rangle \right) }\ \
\ .}
An important consequence of \semib\ is that the width and differential
rates are proportional to the forward matrix element  (between B meson states
of
momentum $P_B$, or more generally between b-hadron states of momentum $P_H$)
of the time-ordered product of the weak current and its hermitean conjugate.
This time-ordered product involves only the quark fields and not the hadron
fields.
The graph that contributes to the physical decay process can  be seen in Fig~3.

\insertfig{Fig~3. Graph contributing to the imaginery part of $T(
j^\dagger_\mu\  j_\nu ) $ .
}{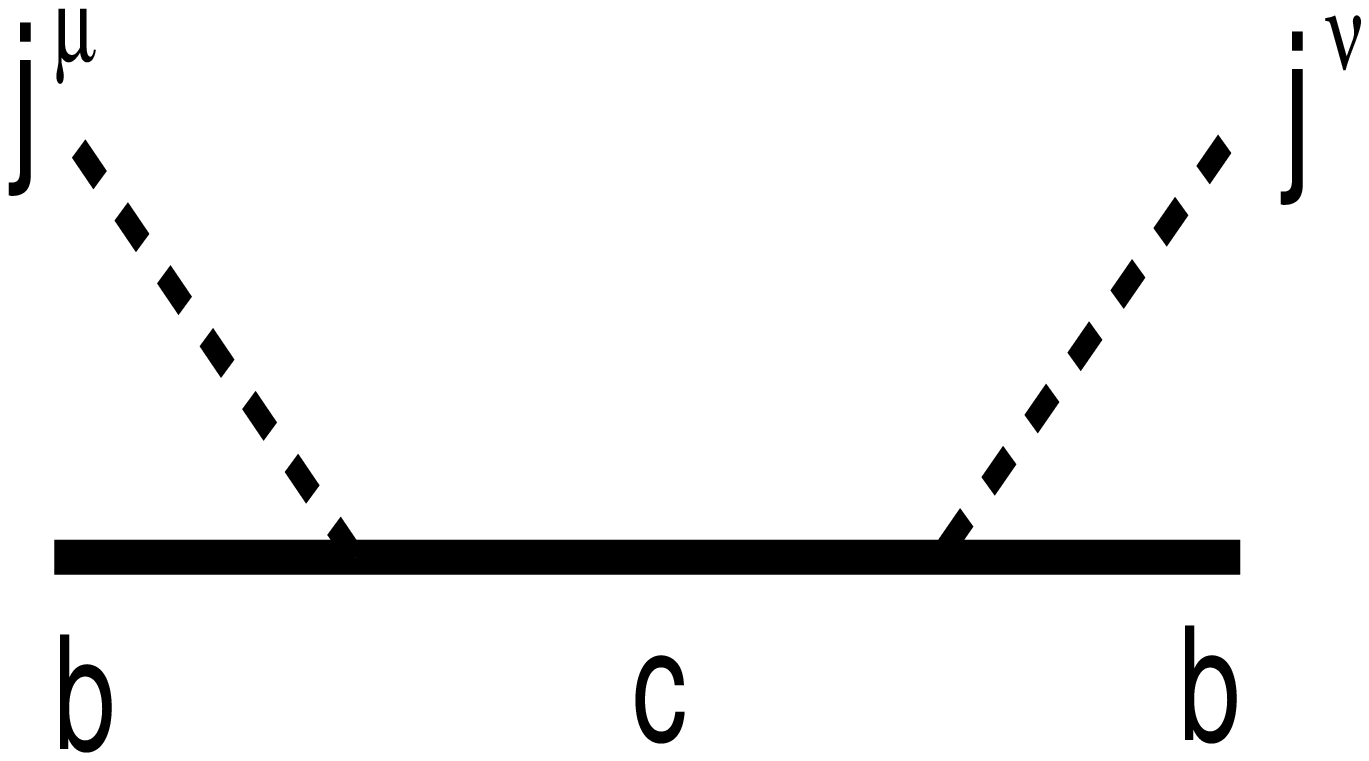}

The integrations over lepton phase space in \semib\ cannot be
computed directly due to the long-distance strong interaction effects
that give rise to incalculable structure.
However, the analyticity of $T( j^\dagger_\mu\  j_\nu )$ allows the
integrations to be deformed
away from the real axis where the incalculable structure lies, into a
region where
it can be computed in perturbative QCD \cgg\ .
This is completely analogous to the methods used in analysing
$e^+e^-\rightarrow {\rm hadrons}$, deep inelastic lepton scattering and
inclusive
semihadronic $\tau$ decays. The appropriate momentum scale at which to
evaluate quantities is at $|s|=m_b$, and hence we have a perturbative expansion
in
$\alpha_s(m_b)$.  We perform
an operator product expansion on the time-ordered product and for
convenience we will match
onto Heavy Quark Fields of fixed velocity $v$.  This provides a
systematic expansion of the
time-ordered product in powers of $\Lambda_{\rm QCD}/m_b$.
Therefore the rate for the inclusive decay \semib\ is a double power
series expansion in
$\alpha_s(m_b)$ and $\Lambda_{\rm QCD}/m_b$.

It was shown in \cgg\ that the leading term in the operator product
expansion reproduces the
parton model result
\eqn\parton{\Gamma^{(0)} = |V_{cb}|^2{G_F^2m_b^5\over
192\pi^3}f(m_c^2/m_b^2)\ \ \ ,}
where $f(x)$ is a function arising from the phase space integral and is
given by
$f(x)=1-8x+8x^3-x^4-12x^2\log x$.   Notice that the mass that appears in
the numerator is the
quark mass and not the meson mass.  More precisely, it is the mass for
which the residual mass in the Heavy Quark Effective Theory (HQET) vanishes.
Further, they showed that
there are no corrections at order $1/m_b$ to the rate.  This is a very
important result as it means that nonperturbative effects arising from the
light
degrees of freedom in the b-hadron
are suppressed by $\left( \Lambda_{\rm QCD}/m_b\right)^2$, and are
expected to be very small.  Further, the corrections at order $1/m_b^2$
are determined by matrix elements of
operators that are process independent.  Neglecting $V_{ub}$ the
expression for the semileptonic inclusive width is
\cgg\bsuva\mwa\mannel\
\eqn\semiinc{\eqalign{
\Gamma (B\rightarrow e^-\overline{\nu_e}+ anything)  =
& |V_{cb}|^2{G_F^2m_b^5\over 192\pi^3}
\left[ \left(
1-{2\alpha_s(\mu_b)\over3\pi}g(m_c^2/m_b^2) + {\lambda_1\over 2 m_b^2}
\right) f(m_c^2/m_b^2) \right.  \cr
&\left.  -{9\lambda_2\over 2 m_b^2} h(m_c^2/m_b^2)
\right] }   \ \ \ \ ,}
where $g(x)$ can be found in
\ref\cm{N. Cabibbo and L. Maiani, \pl{79}{1978}{109}.} and
$h(x)=1-{8\over 3}x-8x^2+8x^3+{5\over 3}x^4+4x^2\log x$.
The quantities $\lambda_{1,2}$ are the nonperturbative matrix elements
given by
\eqn\mat{\eqalign{
2M_B\lambda_1 = &\ \langle\ B\ |\overline{h} (iD)^2 h |\ B\ \rangle\cr
6M_B\lambda_2 = &\ \langle\ B\ |\overline{h} {i\over
2}\sigma^{\mu\nu}G_{\mu\nu}
h |\ B\ \rangle }\ \ \ \ ,} where $\lambda_1$ corresponds to the fermi
motion of the heavy
quark induced by the light degrees of freedom and $\lambda_2$ arises from
the chromomagnetic interaction of the heavy quark with the light degrees of
freedom. The $B^*-B$ mass splitting gives $\lambda_2(m_b)=0.12 {\rm GeV^2}$
(this
is scale dependent and  $\lambda_2(m_c)=0.10 {\rm GeV^2}$ ).
Expressions similar to \semiinc\ can be found for other factorisable
processes such as
$B\rightarrow\gamma+ $hadrons\  \bsuva\fls\
and $B\rightarrow e^+e^-+$ hadrons\   \fls\  involving the same nonperturbative
matrix elements $\lambda_{1,2}$.

One may use \semiinc\ to extract a range of values for $|V_{cb}|$
\ref\ls{M. Luke and M.J. Savage, \pl{321}{1994}{88}.}
\ref\bigiuralta{I.I. Bigi and N.G. Uraltsev, cern-th-7063-93,
hep-ph-9311243.}
once
$\lambda_1$, $m_b$ and the appropriate scale at which to evaluate
$\alpha_s$ are known.  Unfortunately, these quantities are not known at
present and we make an estimate for the range of each in order to proceed.
The mass $m_b$ is the mass for which the residual mass term in HQET
vanishes and we expect that it differs from the B-meson mass by
$\ltap 1 GeV$.  In fact, the quark mass and meson mass are related via
\eqn\mass{\eqalign{M_B = &\ m_b\left( 1 + {\overline{\Lambda}\over m_b} -
{\lambda_1+3\lambda_2\over m_b^2} + ....\right) \cr
M_{B^*} = &\ m_b\left( 1 + {\overline{\Lambda}\over m_b} -
{\lambda_1-\lambda_2\over m_b^2} + ....\right) }\ \ \ \ ,}
where $\lambda_{1,2}$ are the same nonperturbative matrix elements that
appear in \semiinc\  and $\overline{\Lambda}$ is the energy of the light
degrees of
freedom. Secondly, as $\lambda_1$ is a nonperturbative matrix element arising
from
the interactions of the heavy quark with the light degrees of freedom we
expect it is or order $\Lambda_{\rm QCD}$ and also  $\ltap 1 GeV$.
The largest uncertainty in the extraction of $|V_{bc}|$ comes from the
uncertainty
in the scale at which to evaluate  $\alpha_s$ in the
one-loop corrections.  It is reasonable to assume that the scale lies
somewhere between  $m_b/3$ and $m_b$ (suggested by inclusive $\tau$ decays and
$e^+e^-\rightarrow $hadrons) but this can only be resolved with a two loop
calculation.
We find a range for $|V_{bc}|$ as shown by the gray region in
Fig.~4.   The vertical error bar corresponds to the value of $V_{bc}$ extracted
from
exclusive $B$ decays.

\insertfigbig{Fig~4. $|V_{bc}|$ verses $\overline{\Lambda}$ (MeV) for
$\tau_B=1.49$ps.}{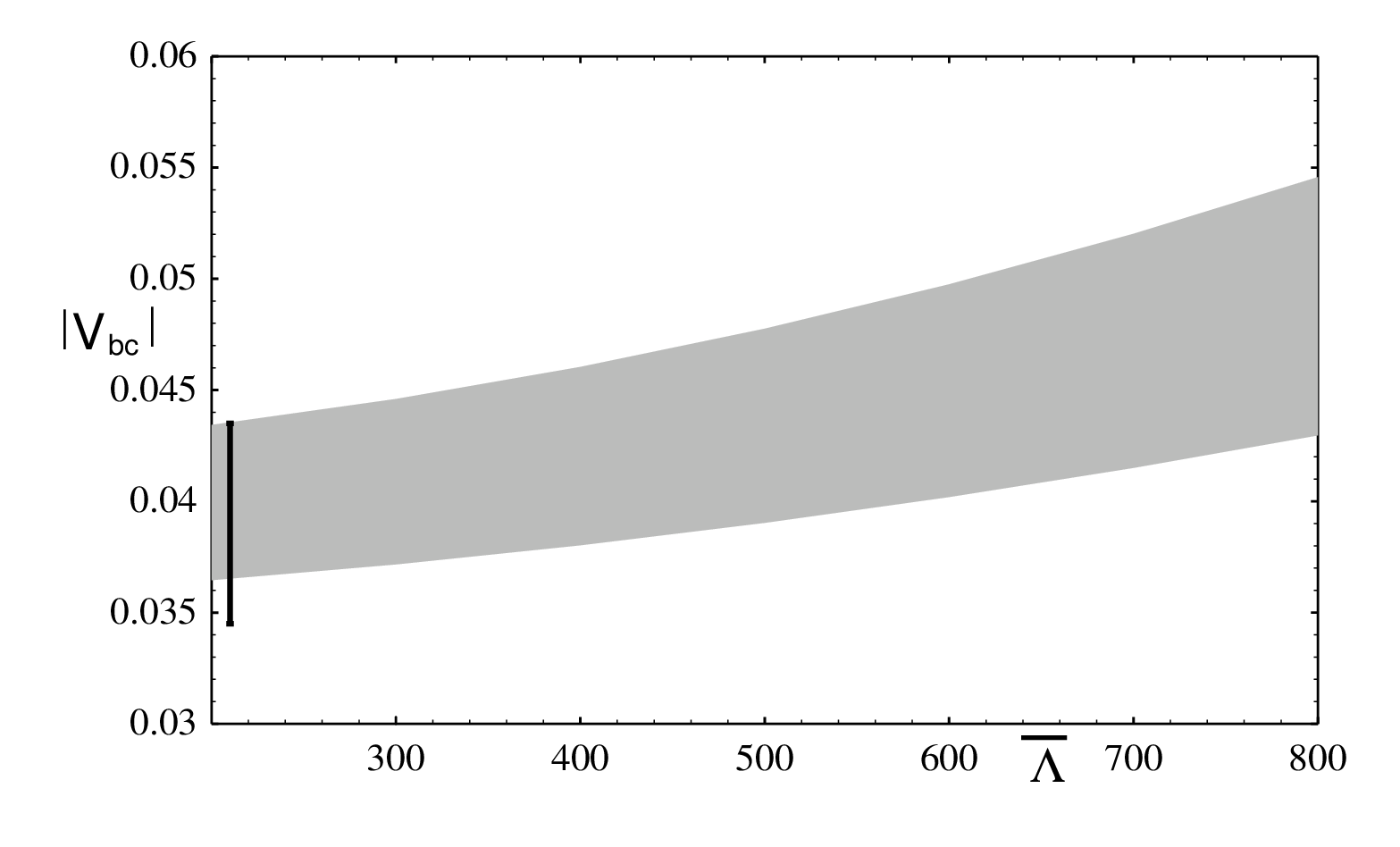}

The inclusive semileptonic decay of charmed hadrons may be addressed
using the same techniques as discussed above.
On one hand, one may be pessimistic about such an approach being applicable to
charmed systems as the mass of the charm quark  is not substantially larger
than
the QCD scale and  higher order corrections are expected to be important.
On the other hand, this feature may be desirable just as it is for
inclusive decays of the $\tau$ lepton.
The significant QCD corrections to the $\tau$ decay rate allows  a precise
determination of $\alpha_s(m_\tau )$ \ref\braat{E. Braaten, S. Narison and
A. Pich, \np{373}{1992}{581}.}.

We can test the inclusive formalism for the charm system\foot{This is
work done in collaboration with M. Luke} by comparing the semileptonic
inclusive
widths for $D^{+,0}$ and  $D_s^+$. Notice that any flavour dependence of  the
inclusive widths can only  arise at order
$1/m_Q^2$ and higher, as the leading term does not depend on the light
degrees of freedom.  Let us write the mass of the charmed mesons as
\eqn\masses{\eqalign{
M_{D} = &\ m_c+ \overline{\Lambda}^u - {\lambda_1^u+3\lambda_2^u\over 2
m_c} + {\cal O}(1/m_c^2)\cr
M_{D^*} = &\ m_c+ \overline{\Lambda}^u - {\lambda_1^u-\lambda_2^u\over 2
m_c}  +  {\cal O}(1/m_c^2)\cr
M_{D_s} = &\ m_c+ \overline{\Lambda}^s - {\lambda_1^s+3\lambda_2^s\over 2
m_c}  +  {\cal O}(1/m_c^2)\cr
M_{D_s^*} = &\ m_c+ \overline{\Lambda}^s - {\lambda_1^s-\lambda_2^s\over
2 m_c}  +  {\cal O}(1/m_c^2)}\ \ \ ,}
where $\lambda_{1,2}^u$ are the matrix elements for mesons with $u$- or
$d$-type light degrees of freedom while $\lambda_{1,2}^s$ are those appropriate
for
$s$-type light  degrees of freedom.
Similar formulae are true for the $B$ and $B_s$ mesons by
simply replacing  $m_c$ with $m_b$.
It is clear that we can determine $\lambda_2^{u,s}$ from the above
expressions and the measured masses, as already discussed above.
We can extract the difference  $\lambda^s-\lambda^u$ from \masses\  by forming
suitable sums and  difference (see also \bigiuralta\ ) of the meson masses.
If we let
$\Sigma_H = M_H + 3M_{H^*}$, then

\eqn\lamsu{\Sigma_B-\Sigma_{B_s}-\Sigma_D+\Sigma_{D_s} =
\left(\lambda_1^u-\lambda_1^s\right)\left({1\over m_c} - {1\over  m_b}\right)\
\ \ ,}
(a similar sum can be formed to determine the difference between $\lambda_1$ in
baryons and mesons \mwa ) and inserting the relevent hadron masses we find that
\eqn\lamnum{\eqalign{
\lambda_1^u-\lambda_1^s = &\  (-4.8\pm 3.2)\times 10^{4} {\rm MeV^2}\cr
\lambda_2^u-\lambda_2^s = &\  (-0.5\pm 2.4)\times 10^{4} {\rm MeV^2}}\ \  \ .}
We note that we have used $M_{B_s^*}-M_{B_s}=47.0\pm 2.6 {\rm MeV}$ as
found from \ref\bstar{ J. Lee-Franzini etal., \pl{65}{1990}{2947}.  }, it is
important that the $B_s^*$ mass be determined with greater precision in order
to
reduce the errors  in \lamnum\ .   We should also note that there are
corrections to
\lamsu\lamnum\ arising at  next order in $1/m_c$
that we have neglected in our analysis.    From these determinations we
can predict the ratio of inclusive decays
\eqn\incsu{\eqalign{
{\Gamma (D_s\rightarrow e\nu+ anything)\over \Gamma (D\rightarrow e\nu+
anything)}
 =&\  \left( 1+ {\lambda_1^s-\lambda_1^u\over 2m_c^2} \right)
- {9\over 2}{\lambda_2^s-\lambda_2^u\over m_c^2}{h(0)\over f(0)} \cr
& \sim 1.00\pm 0.03 }\ \ \ \ .}
This may or may not be a surprise to the reader.  Despite the fact that
exclusive
modes are generally not related by  flavour SU(3) and therefore
one does not naively expect any relation between inclusive widths, we
find that the inclusive formalism leads us to believe that the semileptonic
inclusive
widths  between the $D_s$ and $D$  should be the same up to a few percent.
This is a testable  prediction that will give us
valuable information on the inclusive formalism for charm hadrons.

So far we have only discussed decay rates  and have not examined
differential  spectra resulting from inclusive decays.
These spectra can be obtained from the formalism developed above.  One of
the more interesting applications is to the endpont region of
$B\rightarrow X_u e^-\overline{\nu}_e$ where one hopes to be able to
extract $|V_{ub}|^2$. The inclusive electron spectrum can be computed from
\semib\
and is found to be
\mwa\buv\bksv\
\eqn\diff{{1\over \Gamma}{d\Gamma\over dy} = 2yF(y)\Theta(1-y)
-{\lambda_1+33\lambda_2\over 3m_b^2}\delta(1-y)
-{\lambda_1\over 3m_b^2}\delta^\prime(1-y)\ \ \ ,}
where the normalised electron energy is $y=2E_e/m_b$, the width for this
mode is
$\Gamma~=~|V_{ub}|^2{G_F^2m_b^2\over 192\pi^3}$ and the spectrum inside
the parton model boundaries is
\eqn\nonsing{F(y) = y(3-2y) + y^2{5\lambda_1\over 3m_b^2} +
y(6+5y){\lambda_2\over m_b^2}\ \ \ \ \ .}
The contributions from higher dimension operators are singular at the
parton model endpoint,
in fact some  operators  of order $1/m_b^n$ have singularities up to
$\delta^{(n-1)}(1-y)$.
At first this looks pretty sick but remember that the parton model
kinematic are contained entirely within the corresponding hadron
kinematics, as shown in Fig.~5.

\insertfigsmall{Fig~5.\ Dalitz plot for $B\rightarrow X_ue\nu$.}{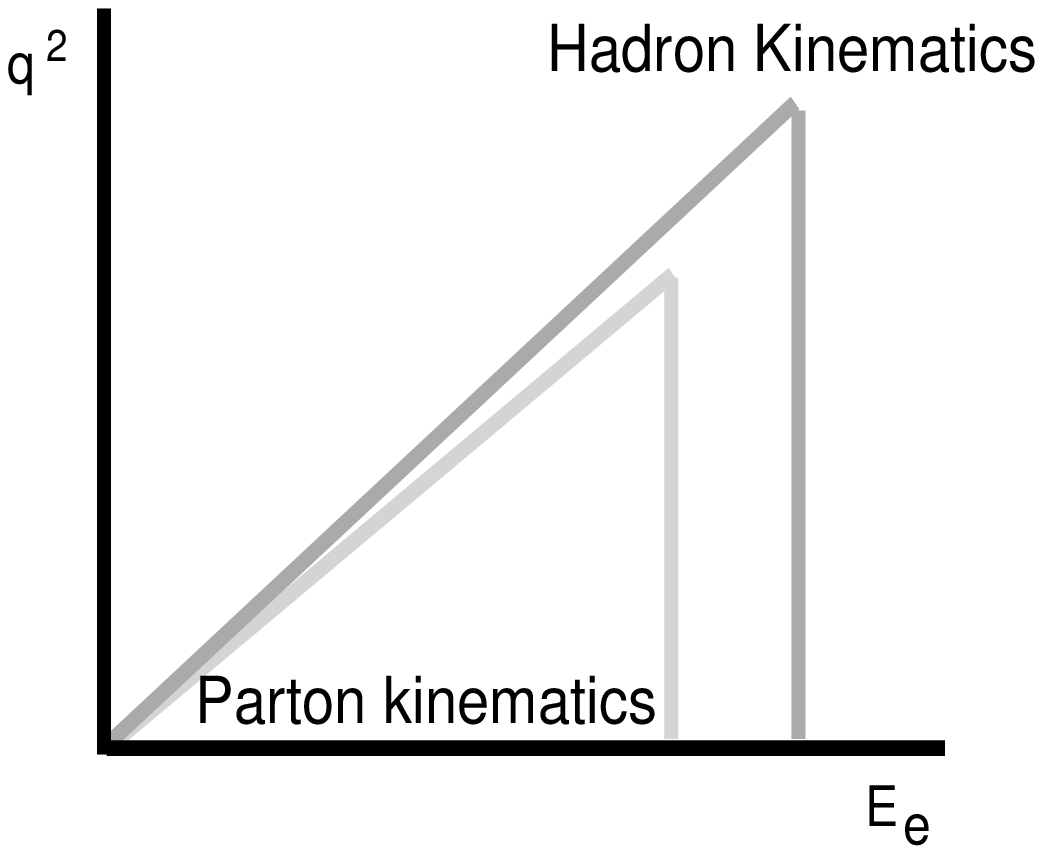}

The effect of this singular behaviour at the parton model endpoint can be
seen by power series expanding a theta function
\ref\neub{M. Neubert, cern-th.7087/93 (1993).}
\eqn\endpt{\Theta(x+\epsilon) = \Theta(x) + \epsilon\delta(x) + {1\over
2}\epsilon^2\delta^\prime(x) + ......\ \ \ .}
It is clear from \diff\  that we cannot absorb the higher order terms
simply by shifting the parton model endpoint.
As demonstrated explicitly in \neub\ we can write the differential rate
as
\eqn\shiftdiff{{1\over \Gamma_{bu}}{d\Gamma\over dy} = 2yF(y)
{1\over N}\sum_i^N\Theta(1-y+\epsilon_i)\ \ \ ,}
from which we see that
\eqn\relhigh{
{1\over N}\sum_i^N \epsilon_i =  -{\lambda_1+33\lambda_2\over 6 m_b^2}
\ \ \ {\rm and }\ \ \
{1\over N}\sum_i^N \epsilon^2_i =  -{\lambda_1\over 3 m_b^2}\ \ \ .}
Setting $\epsilon_i\sim 1/m_b$ we see
that there is a nontrivial cancellation occurring between the
$\epsilon_i$ as there are no
$1/m_b$ corrections to the spectrum or rate ( the sum of terms of order
$1/m_b$ is of order $1/m_b^2$). The physics from this rewriting of the spectrum
is that the
parton model kinematics have spilled over into the region beyond the parton
model but
within the hadron kinematics. The amount is of order $1/m_b$ and as expected is
proportional to the nonperturbative matrix elements.
By resumming the terms singular at the endpoint into a smooth function
we see that the formalism includes the nonperturbative bound state
effects and transforms from parton to hadron kinematics.
It is precisely this region that is important for an
extraction of  $V_{ub}$. Ignoring perturbative strong interaction corrections
one can  show that
the moments of the spectrum beyond the endpoint are related to the
forward matrix elements
of higher dimension operators in the heavy quark expansion, as you may
have already
guessed from  \relhigh\ .
However, the inclusion of perturbative strong interaction
corrections, particularly the $\log^k (1-y)$  that appear near the
endpoint, casts some uncertainty on this method.
One finds that the energy interval for which
such logs become large is much larger than $\Lambda_{\rm QCD}/m_b$
( the region
where the nonperturbative effects are large) as $m_b\rightarrow \infty$
\ref\fjmw{A.F. Falk etal., UCSD/PTH 93-38, (1993).}.
However, for the finite value of
$m_b$ that occurs in nature it is not clear cut one way or the other.
The leading logs can be resummed into the Sudakov formfactor which forces
the rate to
vanish at the endpoint but  the subleading logs need to be summed in
order to make a better guess at the behaviour in this region.

In conclusion, the theory of inclusives decays of hadrons containing
heavy quarks has progressed in leaps and bounds in the last few years.
The parton model decay rate is seen to emerge as the leading term in a
systematic expansion in $\Lambda_{\rm QCD}/m_Q$ and $\alpha_s(m_Q)$
for inclusive decays such
as $\Lambda_b\rightarrow X_ce\nu$, $B\rightarrow X_ce\nu$, $B\rightarrow
X_s\gamma$ and $B\rightarrow X_sl^+l^-$.   An important  feature of the
analysis is that
there are no $1/m_Q$ corrections to the rate, the leading corrections arise
at $1/m_Q^2$ and  $\alpha_s(m_Q)$.
It seems that  extractions of the weak mixing angle $V_{cb}$ from inclusive
decays are competitive with those from exclusives decays $B\rightarrow
D^{(*)}e\nu$.   The largest
uncertainty in the extraction arises from perturbative strong interaction
corrections, in particular the uncertainty in the scale at which to evaluate
$\alpha_s$.
This issue can be resolved by a two loop calculation. There are measurements
that can be
made in the charm sector that determine the size of higher order corrections.
Finally, we
have seen how the parton model kinematics become hadron kinematics.

\bigskip\bigskip
\centerline{\bf Acknowledgements}

I thank Michael Luke for many discussions on the material presented here.
I would also like to thank the organisers of this conference for putting
together a very  stimulating meeting.
This work was supported in part by the Department of Energy under
contract DE--FG02--91ER40682.

\listrefs
\vfill\eject
\bye